\documentclass[preprint,amsmath,amssymb,aps,showkeys,showpacs]{revtex4}
\usepackage[english]{babel}
\usepackage{graphicx}
\usepackage{graphics}
\usepackage{amsmath}
\usepackage{dcolumn}
\usepackage{amssymb}
\usepackage{bm}
\usepackage[latin1]{inputenc}

\begin{document}
\title{He-McKellar-Wilkens effect and Scalar Aharonov-Bohm effect for a neutral particle based on the Lorentz symmetry violation}
\author{K. Bakke}
\email{kbakke@fisica.ufpb.br}
\affiliation{Departamento de F\'isica, Universidade Federal da Para\'iba, Caixa Postal 5008, 58051-970, Jo\~ao Pessoa, PB, Brazil.}

\author{E. O. Silva}
\affiliation{Departamento de F\'{\i}sica, Universidade Federal do Maranh\~{a}o, Campus Universit\'{a}rio do Bacanga, 65085-580, S\~{a}o Lu\'{\i}s, MA, Brazil}

\author{H. Belich}
\affiliation{Departamento de F\'{\i}sica e Qu\'{\i}mica, Universidade Federal do Esp\'{\i}rito Santo, Av. Fernando Ferrari, 514, Goiabeiras, 29060-900, Vit\'{o}ria, ES, Brazil.}

\begin{abstract}
In this contribution, we discuss the He-McKellar-Wilkens effect and the Scalar Aharonov-Bohm effect for neutral particles based on the Lorentz symmetry violation background, by showing that the background of the Lorentz symmetry violation yields abelian quantum phases for a neutral particle. We also study the nonrelativistic bound states for a neutral particle interacting with a Coulomb-like potential based on the Lorentz symmetry violation background given by a fixed vector field parallel to the radial direction.
\end{abstract}

\maketitle

\section{Introduction}

In recent decades, the He-McKellar-Wilkens effect \cite{hm,w} and the Scalar Aharonov-Bohm effect for neutral particles \cite{zei,zei2,anan2} have attracted a great deal of attention in studies of quantum effects that deal with geometric phases \cite{bohm}. Geometric quantum phases is the term denominated by Berry \cite{berry} to describe the phase shift acquired by the wave function of a particle during an adiabatic quantum evolution in a closed path. Geometric phases have been extended to non-adiabatic quantum evolutions in a closed path by Aharonov and Anandan in \cite{ahan}, which is called the Aharonov-Anandan quantum phase. Well-known quantum effects related to geometric phases, corresponding to particular cases of the Aharonov-Anandan quantum phase \cite{ahan}, are the Aharonov-Bohm effect \cite{ab}, the dual Aharonov-Bohm effect \cite{dab}, the Scalar Aharonov-Bohm effect \cite{pesk}, the Aharonov-Casher effect \cite{ac}, the Mashhoon effect \cite{r3}, the He-McKellar-Wilkens effect \cite{hm,w}, and the Scalar Aharonov-Bohm effect for neutral particles \cite{zei,zei2,anan2}. In particular, the He-McKellar-Wilkens effect and the Scalar Aharonov-Bohm effect for neutral particles have been studied in noncommutative quantum mechanics \cite{fur1}, in the presence of topological defects \cite{fur2,bf1,bf8}, and in holonomic quantum computation \cite{bf26}. More discussions about quantum phases for electric dipole moment can be found in \cite{anan,whw,hag,whw2,aud,spa,spa1,spa2,spa3}, and the experimental setup to reproduce the field configuration of the He-McKellar-Wilkens effect can be found in \cite{tka,tka1,tka2,tka3}. Furthermore, the extension of the study of the quantum dynamics of a neutral particle with a permanent electric dipole moment has been made for bound states, where the Landau quantization was proposed in \cite{lin} based on the He-McKellar-Wilkens setup. Thus, the Landau quantization based on the He-McKellar-Wilkens setup has been studied under the influence of topological defects \cite{bf4}, via noninertial effects \cite{b}, and in the relativistic quantum mechanics \cite{bf5,b4}.

Recently, the Aharonov-Casher effect has been studied in the context of the Lorentz symmetry breaking \cite{belich,belich2,belich3}, and the relativistic Anandan quantum phase based on the Lorentz symmetry breaking has been discussed in \cite{bbs2}. Discussions about symmetry breaking are well-known in nonrelativistic quantum systems involving phase transitions such as ferromagnetic systems, where the rotation symmetry is broken when the system is under the influence of a magnetic field. For relativistic systems, the study of symmetry breaking can be extended by considering a background given by a $4$-vector field that breaks the symmetry $\mathcal{SO}\left(1,3\right)$ instead of the symmetry $\mathcal{SO}\left(3\right)$. This line of research is known in the literature as the spontaneous violation of the Lorentz symmetry \cite{extra3,extra1,extra2}. This new possibility of spontaneous violation was first suggested in 1989 in a work of Kostelecky and Samuel \cite{extra3}. Kostelecky and Samuel \cite{extra3} indicated that, in the string field theory, the spontaneous violation of symmetry by a scalar field could be extended. This extension has as immediate consequence: a spontaneous breaking of Lorentz symmetry. In the electroweak theory, a scalar field acquires a nonzero vacuum expectation value which yields mass to gauge bosons (Higgs Mechanism). Similarly, in the string field theory, this scalar field can be extended to a tensor field. Nowadays, these theories are encompassed in the framework of the Extended Standard Model (SME) \cite{col} as a possible extension of the minimal Standard Model of the fundamental interactions. For instance, the violation of the Lorentz symmetry is implemented in the fermion section of the Extended Standard Model by two CPT-odd terms: $v_{\mu }\overline{\psi }\gamma ^{\mu }\psi$ and $b_{\mu }\overline{\psi }\gamma_{5}\gamma ^{\mu }\psi$, where $v_{\mu} $ and $b_{\mu}$ correspond to the Lorentz-violating backgrounds \cite{belich,belic,belich2,belich3}. The modified Dirac theory has already been examined in the literature \cite{Hamilton}, and, in the nonrelativistic limit of the modified Dirac theory, the spectrum of energy of the hydrogen atom has been discussed in \cite{Manojr,Nonmini}. Hence, there is an extensive amount of work looking at Lorentz violation, and numerous experimental bounds exist \cite{extra2}. How does the interaction introduced here fits within that context? First we note that the description of a neutral particle immersed in an environment with nonminimal coupling (via Dirac equation) has been little explored. In this work, we intend to investigate possible measurements of geometric phases coming from this violation background, out of the proposal of the SME. We present a new phase that could detect the presence of a background field through interference experiments. Then, we could verify if the Lorentz violation describes neutrino oscillations, which makes of these measurements sensitive probes of new physics \cite{proc}. We would like to point out that the criterion used in this work to suggest the violation of Lorentz symmetry is that the gauge symmetry should be preserved. Other studies of the influence of a symmetry breaking background have been made for bound states for a relativistic neutral particle with an analogue of the permanent magnetic dipole moment in the Lorentz symmetry violation background \cite{bbs}, and for a charged particle describing a circular path in presence of a Lorentz-violating background nonminimally coupled to a spinor and a gauge field \cite{mano}.

In this paper, we discuss the He-McKellar-Wilkens effect \cite{hm,w} and the Scalar Aharonov-Bohm effect for neutral particles \cite{zei,zei2,anan2} based on the Lorentz symmetry violation background, and show that the background of the Lorentz symmetry violation yields a abelian quantum phases for a neutral particle differing from the non-abelian phases of the He-McKellar-Wilkens effect \cite{hm,w} and the Scalar Aharonov-Bohm effect for neutral particles \cite{zei,zei2,anan2}. Moreover, we study the nonrelativistic bound states for a neutral particle interacting with a Coulomb-like potential based on the Lorentz symmetry violation background given by a fixed vector parallel to the plane of motion of the quantum particle. The structure of this paper is: in section II, we discuss the analogues effects of the He-McKellar-Wilkens effect \cite{hm,w} and the Scalar Aharonov-Bohm effect for neutral particles \cite{zei,zei2,anan2} based on the Lorentz symmetry violation background; in section III, we consider a fixed vector parallel to the plane of motion of the quantum particle and discuss the bound states for a neutral particle in a Coulomb-like potential based on the Lorentz symmetry violation background, in section IV, we present our conclusions.

\section{He-McKellar-Wilkens effect and Scalar Aharonov-Bohm effect based on the Lorentz symmetry violation background}

In this section, we discuss the Anandan quantum phase, the He-McKellar-Wilkens effect, and the Scalar Aharonov-Bohm effect for a neutral particle based on the Lorentz symmetry violation background. We start by introducing a nonminimal coupling into the Dirac equation given by
\begin{eqnarray}
i\gamma^{\mu}\partial_{\mu}\rightarrow i\gamma^{\mu}\partial_{\mu}-g\,b^{\mu}\,F_{\mu\nu}\left(x\right)\,\gamma^{\nu},
\label{1}
\end{eqnarray}
where $g$ is a constant, $b^{\mu}$ corresponds to a fixed $4$-vector that acts on a vector field breaking the Lorentz symmetry violation, the tensor $F_{\mu\nu}\left(x\right)$ corresponds to the electromagnetic tensor, whose components are $F_{0i}=-F_{i0}=-E_{i}$, and $F_{ij}=-F_{ji}=\epsilon_{ijk}B^{k}$. This coupling follows a different proposal of SMS \cite{belich,belic,belich2,belich3}. The novelty is that instead of using the dual electromagnetic tensor, we put the own field strength tensor. The $\gamma^{\mu}$ matrices are defined in the Minkowski spacetime in the form \cite{greiner}:
\begin{eqnarray}
\gamma^{0}=\hat{\beta}=\left(
\begin{array}{cc}
1 & 0 \\
0 & -1 \\
\end{array}\right);\,\,\,\,\,\,
\gamma^{i}=\hat{\beta}\,\hat{\alpha}^{i}=\left(
\begin{array}{cc}
 0 & \sigma^{i} \\
-\sigma^{i} & 0 \\
\end{array}\right);\,\,\,\,\,\,\Sigma^{i}=\left(
\begin{array}{cc}
\sigma^{i} & 0 \\
0 & \sigma^{i} \\	
\end{array}\right),
\label{2}
\end{eqnarray}
with $\vec{\Sigma}$ being the spin vector. The matrices $\sigma^{i}$ are the Pauli matrices, and satisfy the relation $\left(\sigma^{i}\,\sigma^{j}+\sigma^{j}\,\sigma^{i}\right)=2\eta^{ij}$. Thus, in the Lorentz symmetry violation background described by the nonminimal coupling (\ref{1}), the Dirac equation becomes
\begin{eqnarray}
i\frac{\partial\psi}{\partial t}=m\hat{\beta}\psi+\vec{\alpha}\cdot\vec{p}\,\psi-g\,b^{0}\vec{\alpha}\cdot\vec{E}\psi-g\,\vec{\alpha}\cdot\left(\vec{b}\times\vec{B}\right)\psi+g\,\vec{b}\cdot\vec{E}\psi.
\label{3}
\end{eqnarray}

By observing the nonminimal coupling introduced in Eq. (\ref{1}), and the Dirac equation (\ref{3}), one can observe that this kind of coupling is a new possibility of a Lorentz symmetry breaking mechanism which opens a new possibility to detect the violation of the Lorentz symmetry, for instance, through the appearance of geometric phases in the wave function of a neutral particle. The motivation for studying this type of violation is the fact that a fundamental theory that incorporates the four fundamental forces of nature, where in the range of low energy, it appears correcting the standard model with new terms. The Lorentz violation terms are generated as vacuum expectation values of tensors defined in a high energy scale. We propose the investigation of nonminimal coupling terms in the context of Lorentz-violating models involving some fixed background, gauge fields, and fermion fields. The main purpose is to figure out whether such new couplings are able to induce geometric quantum phases in the matter sector.

In this work, we are interested in discussing the nonrelativistic behavior of the spin-half neutral particle in a background of the Lorentz symmetry violation described by (\ref{1}). We can obtain the nonrelativistic dynamics of the neutral particle by writing the solution of the Dirac equation (\ref{3}) in the form
\begin{eqnarray}
\psi=e^{-imt}\,\left(
\begin{array}{c}
\phi\\
\chi\\	
\end{array}\right),
\label{4}
\end{eqnarray}
where $\phi$ and $\chi$ are two-spinors, and we consider $\phi$ being the ``large'' component and $\chi$ being the ``small'' component \cite{greiner}. Substituting (\ref{4}) into the Dirac equation (\ref{3}), we obtain two coupled equations of $\phi$ and $\chi$. The first coupled equation is
\begin{eqnarray}
i\frac{\partial\phi}{\partial t}-g\,\vec{b}\cdot\vec{E}\phi=\left[\vec{\sigma}\cdot\vec{p}-g\,b^{0}\vec{\sigma}\cdot\vec{E}-g\,\vec{\sigma}\cdot\left(\vec{b}\times\vec{B}\right)\right]\chi,
\label{5}
\end{eqnarray}
while the second coupled equation is
\begin{eqnarray}
i\frac{\partial\chi}{\partial t}-g\,\vec{b}\cdot\vec{E}\phi+2m\chi=\left[\vec{\sigma}\cdot\vec{p}-g\,b^{0}\vec{\sigma}\cdot\vec{E}-g\,\vec{\sigma}\cdot\left(\vec{b}\times\vec{B}\right)\right]\phi.
\label{6}
\end{eqnarray}
With $\chi$ being the ``small" component of the wave function, we can consider $\left|2m\chi\right|>>\left|i\frac{\partial\chi}{\partial t}\right|$, and $\left|2m\chi\right|>>\left|g\,\vec{b}\cdot\vec{E}\phi\right|$, thus, we can write
\begin{eqnarray}
\chi\approx\frac{1}{2m}\left[\vec{\sigma}\cdot\vec{p}-g\,b^{0}\vec{\sigma}\cdot\vec{E}-g\,\vec{\sigma}\cdot\left(\vec{b}\times\vec{B}\right)\right]\phi.
\label{7}
\end{eqnarray}
Substituting (\ref{7}) into (\ref{5}), we obtain the Schr\"odinger-Pauli equation
\begin{eqnarray}
i\frac{\partial\phi}{\partial t}=\frac{1}{2m}\left[\vec{p}-g\,b^{0}\,\vec{E}-g\left(\vec{b}\times\vec{B}\right)\right]^{2}\phi+g\,\vec{b}\cdot\vec{E}\phi-\frac{1}{2m}\,\vec{\sigma}\cdot\vec{B}_{\mathrm{eff}}\,\phi,
\label{8}
\end{eqnarray}
where we have defined an effective magnetic field given by
\begin{eqnarray}
\vec{B}_{\mathrm{eff}}=\vec{\nabla}\times\left[g\,b^{0}\vec{E}+g\,\left(\vec{b}\times\vec{B}\right)\right].
\label{9}
\end{eqnarray}

By considering a space-like vector given by $b^{\mu}=\left(0,0,0,b^{3}\right)$, we can consider an analogue of the permanent electric dipole moment of a neutral particle in the form: $\vec{d}=g\vec{b}=g\left(0,0,b^{3}\right)$. Hence, by applying the Dirac phase factor method \cite{dirac}, we can write the wave function of the neutral particle in the form $\phi=e^{i\Phi}\,\phi_{0}$, where $\phi_{0}$ is the solution of the following equation
\begin{eqnarray}
i\frac{\partial\phi_{0}}{\partial t}=-\frac{1}{2m}\,\nabla^{2}\,\phi_{0}-\frac{1}{2m}\,\vec{\sigma}\cdot\vec{B}_{\mathrm{eff}}\,\phi_{0}.
\label{9.1}
\end{eqnarray}
Thus, the Anandan quantum phase \cite{anan} for a neutral particle with a permanent electric dipole moment based on the Lorentz symmetry violation background is given by
\begin{eqnarray}
\Phi_{\mathrm{A}}=g\oint\left(\vec{b}\times\vec{B}\right)_{\mu}\,dx^{\mu}-g\int^{\tau}_{0}\vec{b}\cdot\vec{E}\,dt.
\label{10}
\end{eqnarray}

By comparing with the Anandan quantum phase for a neutral particle with a permanent electric dipole moment given in Ref. \cite{anan2}, we can see that the Anandan quantum phase (\ref{10}) differs from that one of Ref. (\cite{anan2}) because the geometric phase (\ref{10}) is an abelian quantum phase, while the geometric phase in \cite{anan2} is a non-abelian quantum phase. This difference results from the Lorentz symmetry violation background given by a fixed vector in (\ref{10}).

An special case occurs when we consider a magnetic field produced by a linear distribution of magnetic charges on the $z$-axis. This linear distribution produces a radial magnetic field given by $\vec{B}=\frac{\lambda_{m}}{\rho}\hat{\rho}$, where $\rho^{2}=x^{2}+y^{2}$ \cite{hm,w}. In this way, by applying the Dirac phase factor method \cite{dirac}, we have that $\psi_{0}$ is the solution of the following equation
\begin{eqnarray}
i\frac{\partial\phi_{0}}{\partial t}=-\frac{1}{2m}\,\nabla^{2}\,\phi_{0},
\label{11}
\end{eqnarray}
and the geometric phase acquired by the wave function of the neutral particle is
\begin{eqnarray}
\Phi_{\mathrm{HMW}}=g\oint\left(\vec{b}\times\vec{B}\right)\cdot d\vec{r}=2\pi\,g\lambda_{m}\,b^{3},
\label{12}
\end{eqnarray}
which corresponds to the analogue effect of the He-McKellar-Wilkens effect \cite{hm,w} based on the violation of the Lorentz symmetry. Note that the analogue of the He-McKellar-Wilkens effect in (\ref{12}) is given by an abelian quantum phase, which differs from the non-abelian geometric phase obtained in \cite{hm,w}. This results from the Lorentz symmetry violation background given by a fixed vector. Moreover, the abelian geometric phase (\ref{12}) does not depend on the velocity of the neutral particle, that is, the geometric phase is nondispersive \cite{disp,disp2,disp3} in the same way of that one of the He-McKellar-Wilkens effect \cite{hm,w}.

Another special case of the Anandan quantum phase (\ref{10}) occurs when we consider a uniform electric field along the $z$-axis, $\vec{E}=E_{0}\hat{z}$, where $E_{0}$ is a constant. In this case, by applying the Dirac phase factor method \cite{dirac}, we have that $\phi_{0}$ remains the solution of the Eq. (\ref{11}), but the geometric phase acquired by the wave function of the neutral particle is
\begin{eqnarray}
\Phi_{\mathrm{SAB}}=-g\int^{\tau}_{0}\vec{b}\cdot\vec{E}\,dt=-g\,E_{0}\,\tau\,b^{3},
\label{13}
\end{eqnarray}
which corresponds to the analogue effect of the Scalar Aharonov-Bohm effect for neutral particles \cite{anan2} based on the Lorentz symmetry violation background. We can also note that the analogue of the Scalar Aharonov-Bohm effect for a neutral particle (\ref{13}) is given by an abelian quantum phase, and differs from the non-abelian geometric phase obtained in \cite{anan2}. Again, the appearance of a abelian geometric phase results from the Lorentz symmetry violation background given by a fixed vector. However, in the same way of Scalar Aharonov-Bohm effect for a neutral particle (\ref{13}), the the abelian geometric phase (\ref{13}) does not depend on the velocity of the neutral particle, that is, the geometric phase (\ref{13}) is also nondispersive \cite{disp,disp2,disp3}.

Recently, the Scalar Aharonov-Bohm effect for neutral particles with permanent electric dipole moment has been obtained in the presence of a disclination, where it has been shown that one-qubit quantum gates can be implemented based on the quantum holonomies provided by the non-Abelian geometric phase of the Scalar Aharonov-Bohm effect \cite{bf26}. However, the Scalar Aharonov-Bohm effect for neutral particles given in (\ref{13}) is given by an abelian quantum phase, which cannot be used in studies of quantum holonomies in the same way of \cite{bf26}, but this kind of geometric phase should be interesting in studies of quantum evolutions for abelian anyons \cite{abelian}.

\section{Neutral particle in a Coulomb-like potential based on the Lorentz symmetry violation background}

In this section, we discuss the arising of bound states in the nonrelativistic quantum dynamics of a neutral particle from the influence of the Lorentz symmetry violation background analogous to the Coulomb potential. We begin this section by writing the Dirac equation in curvilinear coordinates. In curvilinear coordinates, for instance, in cylindrical coordinates, the line element of the Minkowski spacetime is writing in the form: $ds^{2}=-dt^{2}+d\rho^{2}+\rho^{2}d\varphi^{2}+dz^{2}$. Thus, by applying a coordinate transformation $\frac{\partial}{\partial x^{\mu}}=\frac{\partial \bar{x}^{\nu}}{\partial x^{\mu}}\,\frac{\partial}{\partial\bar{x}^{\nu}}$, and a unitary transformation on the wave function $\psi\left(x\right)=U\,\psi'\left(\bar{x}\right)$, the Dirac equation can be written in any orthogonal system in the following form \cite{schu}:
\begin{eqnarray}
i\,\gamma^{\mu}\,D_{\mu}\,\psi+\frac{i}{2}\,\sum_{k=1}^{3}\,\gamma^{k}\,\left[D_{k}\,\ln\left(\frac{h_{1}\,h_{2}\,h_{3}}{h_{k}}\right)\right]\psi=m\psi,
\label{2.1}
\end{eqnarray}
where $D_{\mu}=\frac{1}{h_{\mu}}\,\partial_{\mu}$ is the derivative of the corresponding coordinate system, and the parameter $h_{k}$ correspond to the scale factors of this coordinate system. For instance, in cylindrical coordinates, the scale factors are $h_{0}=1$, $h_{1}=1$, $h_{2}=\rho$, and $h_{3}=1$. In this way, the second term in (\ref{2.1}) gives rise to a term called the spinorial connection \cite{schu,b4,bbs,bbs2,bd,weinberg}. By introducing the nonminimal coupling (\ref{1}), the Dirac equation (\ref{2.1}) becomes
\begin{eqnarray}
i\,\gamma^{\mu}\,D_{\mu}\,\psi+\frac{i}{2}\,\sum_{k=1}^{3}\,\gamma^{k}\,\left[D_{k}\,\ln\left(\frac{h_{1}\,h_{2}\,h_{3}}{h_{k}}\right)\right]\psi-g\,b^{\mu}\,F_{\mu\nu}\left(x\right)\,\gamma^{\nu}=m\psi.
\label{2.1a}
\end{eqnarray}
Thus, following the same steps from Eq. (\ref{4}) to (\ref{8}) to obtain the nonrelativistic limit of the Dirac equation (\ref{2.1}), thus, the Schr\"odinger-Pauli equation becomes
\begin{eqnarray}
i\frac{\partial\phi}{\partial t}=\frac{1}{2m}\left[\vec{p}-i\vec{\xi}-g\,b^{0}\,\vec{E}-g\left(\vec{b}\times\vec{B}\right)\right]^{2}\phi+g\,\vec{b}\cdot\vec{E}\phi-\frac{1}{2m}\,\vec{\sigma}\cdot\vec{B}_{\mathrm{eff}}\,\phi,
\label{2.2}
\end{eqnarray}
where we have defined the vector $\vec{\xi}$ in such a way that its components are $-i\xi_{i}=\frac{1}{4}\,\omega_{i ab}\left(x\right)\,\Sigma^{ab}$. Note that, the vector $\vec{\xi}$ in (\ref{2.2}) corresponds to the contribution from the spinorial connection \cite{bbs,bbs2}. We should take into account that there is no contribution to the geometric phases obtained in the last section from the spinorial connection, thus, for this reason, we have not included this term into the Dirac equation (\ref{3}).

Now, we consider an electric field produced by a linear distribution of electric charges on the $z$-axis $\vec{E}=\frac{\lambda}{\rho}\,\hat{\rho}$ (with $\lambda$ being a linear electric charge density, and $\rho=\sqrt{x^{2}+y^{2}}$), and a space-like vector field given by $\vec{b}=b\,\hat{\rho}$, where $b$ is a constant and $\hat{\rho}$ is a unit vector on the radial direction. This electric field configuration describes a possible scenario for a measurement of the Lorentz violation. Thus, we have
\begin{eqnarray}
\vec{b}\cdot\vec{E}=\frac{b\lambda}{\rho},
\label{2.3}
\end{eqnarray}
Note that we have chosen a space-like vector field $\vec{b}$ in (\ref{2.3}) that differs from the choice made in the previous section in the study of geometric phases. We also note that this choice of the external field and the fixed vector makes that $\vec{B}_{\mathrm{eff}}=0$. Hence, the Schr\"odinger-Pauli equation (\ref{2.2}) becomes
\begin{eqnarray}
i\frac{\partial\phi}{\partial t}=-\frac{1}{2m}\left[\frac{\partial^{2}\phi}{\partial\rho^{2}}+\frac{1}{\rho}\frac{\partial\phi}{\partial\rho}+\frac{1}{\rho^{2}}\frac{\partial^{2}\phi}{\partial\varphi^{2}}+\frac{\partial^{2}\phi}{\partial z^{2}}\right]+\frac{1}{2m}\frac{i\sigma^{3}}{\rho^{2}}\frac{\partial\phi}{\partial\varphi}+\frac{1}{8m\rho^{2}}\,\phi+\frac{gb\lambda}{\rho}\,\phi.
\label{2.4}
\end{eqnarray}

We can note that $\phi$ is an eigenfunction of $\sigma^{3}$, whose eigenvalues are $s=\pm1$. Thus, we can write $\sigma^{3}\phi_{s}=\pm\phi_{s}=s\phi_{s}$. Since the operators $\hat{J}_{z}=-i\partial_{\varphi}$ \footnote{It has been shown in Ref. \cite{schu} that the $z$-component of the total angular momentum in cylindrical coordinates is given by $\hat{J}_{z}=-i\partial_{\varphi}$, where the eigenvalues are $\mu=l\pm\frac{1}{2}$.} and $\hat{p}_{z}=-i\partial_{z}$ commute with the Hamiltonian of the right-hand side of (\ref{2.4}), we can take the solutions of (\ref{2.4}) in the form:
\begin{eqnarray}
\phi_{s}=e^{-i\mathcal{E}t}\,e^{i\left(l+\frac{1}{2}\right)\varphi}\,e^{ikz}\,\left(
\begin{array}{c}
R_{+}\left(\rho\right)\\
R_{-}\left(\rho\right)\\	
\end{array}\right),
\label{2.5}
\end{eqnarray}
where $l=0,\pm1,\pm2,\ldots$ and $k$ is a constant. Substituting the solution (\ref{2.5}) into the Schr\"odinger-Pauli equation (\ref{2.4}), we obtain two non-coupled equations for $R_{+}$ and $R_{-}$. After some calculations, we can write the noncoupled equations for $R_{+}$ and $R_{-}$ in the following compact form:
\begin{eqnarray}
R_{s}''+\frac{1}{\rho}R_{s}'-\frac{\zeta_{s}^{2}}{\rho^{2}}R_{s}+\frac{\delta}{\rho}R_{s}+\kappa^{2}\,R_{s}=0,
\label{2.6}
\end{eqnarray}
with $\sigma^{3}R_{s}=\pm R_{s}=sR_{s}$, and where we have defined the following parameters:
\begin{eqnarray}
\zeta_{\pm}&=&\zeta_{s}=l+\frac{1}{2}\left(1-s\right);\nonumber\\
\kappa^{2}&=&2m\mathcal{E}-k^{2};\\
\delta&=&2mgb\lambda.\nonumber
\label{2.7}
\end{eqnarray}

Let us observe the asymptotic behavior of Eq. (\ref{2.6}). When $\rho\rightarrow\infty$, we have
\begin{eqnarray}
\frac{d^{2}R_{s}}{d\rho^{2}}+\kappa^{2}\,R_{s}=0,
\label{2.7a}
\end{eqnarray}
thus, we can find either scattering states $R_{s}\cong e^{i\kappa\rho}$, or if we consider $\kappa^{2}=-\tau^{2}$, we can find bound states, where $R_{s}\cong e^{-\tau\rho}$ \cite{mello,b3}. In this work, we are interested in obtaining the energy levels of bound states, thus, we rewrite the second order differential equation (\ref{2.6}) as
\begin{eqnarray}
R_{s}''+\frac{1}{\rho}R_{s}'-\frac{\zeta_{s}^{2}}{\rho^{2}}R_{s}+\frac{\delta}{\rho}R_{s}-\tau^{2}\,R_{s}=0.
\label{2.8}
\end{eqnarray}
In the following, we make a change of variables given by $\eta=2\tau\rho$, and rewrite (\ref{2.8}) in the form:
\begin{eqnarray}
R_{s}''+\frac{1}{\eta}R_{s}'-\frac{\zeta_{s}^{2}}{\eta^{2}}R_{s}+\frac{\delta}{2\tau\eta}R_{s}-\frac{1}{4}\,R_{s}=0.
\label{2.9}
\end{eqnarray}
In order to get a regular solution at the origin, thus, the solution of the second order differential equation (\ref{2.9}) is given by
\begin{eqnarray}
R_{s}\left(\eta\right)=e^{-\frac{\eta}{2}}\,\eta^{\left|\zeta_{s}\right|}\,F_{s}\left(\eta\right).
\label{2.10}
\end{eqnarray}
Thus, substituting the solution (\ref{2.10}) into the second order differential equation (\ref{2.9}), we obtain that the function $F_{s}\left(\eta\right)$ is a solution of the differential equation:
\begin{eqnarray}
\eta\,F_{s}''+\left[2\left|\zeta_{s}\right|+1-\eta\right]F_{s}'+\left[\frac{\delta}{2\tau}-\left|\zeta_{s}\right|-\frac{1}{2}\right]F_{s}=0.
\label{2.11}
\end{eqnarray}

The second order differential equation (\ref{2.11}) is known in the literature as the confluent hypergeometric equation or the Kummer equation \cite{abra}. One of the solutions of (\ref{2.11}) is called the Kummer function of first kind, which is given by $F_{s}\left(\eta\right)=F\left[\left|\zeta_{s}\right|+\frac{1}{2}-\frac{\delta}{2\tau},2\left|\zeta_{s}\right|+1,\eta\right]$ \cite{abra}. In order to obtain a regular solution of the second order differential equation (\ref{2.11}) at the origin, we must impose the condition where the confluent hypergeometric series becomes a polynomial of degree $n$ (where $n=0,1,2,\ldots$). This occurs when
\begin{eqnarray}
\left|\zeta_{s}\right|+\frac{1}{2}-\frac{\delta}{2\tau}=-n.
\label{2.12}
\end{eqnarray}
This condition makes the radial wave function of the neutral particle to be finite everywhere \cite{landau}. Thus, substituting the parameters defined in (\ref{2.7}), with $\tau^{2}=2m\mathcal{E}-k^{2}$, we obtain
\begin{eqnarray}
\mathcal{E}_{n,\,l}=\frac{2m\left(gb\lambda\right)^{2}}{\left[n+\left|\zeta_{s}\right|+\frac{1}{2}\right]^{2}}+\frac{k^{2}}{2m}.
\label{2.13}
\end{eqnarray}

Hence, the expression (\ref{2.13}) corresponds to the energy levels of the bound states in the nonrelativistic quantum dynamics of a neutral particle which arises from a Coulomb-like potential based on the Lorentz symmetry violation background, that is, based on the interaction of a fixed vector parallel to plane of motion of the neutral particle and a radial electric field produced by a linear distribution of electric charges on the $z$-axis.

\section{conclusions}

We have discussed the Anandan quantum phase, the He-McKellar-Wilkens effect, the Scalar Aharonov-Bohm effect, and the interaction between a neutral particle with a Coulomb-like potential based on the Lorentz symmetry violation background. This study has been made by introducing a nonminimal coupling into the Dirac equation, and followed by the procedure in taking the nonrelativistic limit of the Dirac equation. Thus, by considering a Lorentz symmetry violation background given by a space-like vector $b^{\mu}=\left(0,0,0,b^{3}\right)$, we have assumed an analogue of the permanent electric dipole moment of a neutral particle given by $\vec{d}=g\vec{b}=g\left(0,0,b^{3}\right)$, and obtained the general form for the geometric quantum phase for a neutral particle which is called the Anandan quantum phase. As a special case of the Anandan quantum phase, we have considered the interaction between this analogue of the permanent electric dipole moment with a radial magnetic field produced by a linear distribution of magnetic charges, and as a result of this interaction, we have obtained the analogue effect of the He-McKellar-Wilkens effect based on the Lorentz symmetry violation. Another special case of the Anandan quantum phase has been obtained by considering the interaction between a uniform electric field along the $z$ direction with the analogue of the permanent electric dipole moment. This interaction gives rise to the the Scalar Aharonov-Bohm effect for neutral particles based on the Lorentz symmetry violation background.

Finally, we have discussed the quantum dynamics of a neutral particle under the influence of a radial electric field produced by a linear distribution of electric charges on the $z$ direction based on the Lorentz symmetry violation background given by a space-like vector being parallel to the radial direction. We have seen that the term $g\,\vec{b}\cdot\vec{E}$ of the Schr\"odinger-Pauli equation (\ref{2.2}) gives rise to a Coulomb-like potential, where the nonrelativistic energy levels of the bound states depend on the linear charge density, and the parameter associated to the Lorentz symmetry violation background.  We should note that the problem of detecting such a violation at low energies is that the expected value in terms of vacuum is weak. Thus, what has been done is to establish limits of energy in which there cannot be seen this breaking \cite{belich,belich2,belich3}.

This work has been supported by the Brazilian agencies CNPq, CAPES/PNPD, and FAPEMA.


\begin{thebibliography}{99}

\bibitem{hm} X. G. He and B. H. J. McKellar, Phys. Rev. A {\bf47}, 3424 (1993).

\bibitem{w} M. Wilkens, Phys. Rev. Lett. {\bf72}, 5 (1994).


\bibitem{zei}  A. Zeilinger, J. Phys. Colloq. (France) {\bf45}, C3-213 (1984).

\bibitem{zei2} A. Zeilinger, in {\it Fundamental Aspects of Quantum Theory}, edited by V. Gorini and A. Frigero (Plenum, New York, 1985).

\bibitem{anan2} J. Anandan, Phys. Rev. Lett. {\bf85}, 1354 (2000).

\bibitem{bohm} A. Bohm, A. Mostafazadeh, H. Koizumi, Q. Niu and J. Zwanziger, \textit{The geometric phase in quantum systems: foundations, mathematical concepts and applications in molecular and condensed matter physics} (Springer-Verlag, Berlin, Heidelberg, 2003).

\bibitem{berry} M. V. Berry, Proc. R. Soc. Lond. A {\bf392}, 45 (1984).


\bibitem{ahan} Y. Aharonov and J. Anandan, Phys. Rev. Lett. {\bf 58}, 1593 (1987).


\bibitem{ab} Y. Aharonov and D. Bohm,  Phys. Rev. {\bf115}, 485 (1959).

\bibitem{dab} J. P. Dowling, C. Williams and J. D. Franson, Phys. Rev. Lett. {\bf83}, 2486 (1999); C. Furtado and G. Duarte, Phys. Scr. {\bf71}, 7 (2005).

\bibitem{pesk} M. Peshkin and A. Tonomura, \textit{The Aharonov-Bohm Effect} (Springer-Verlag, in: Lecture Notes in Physics, Vol. 340, Berlin, 1989).



\bibitem{ac} Y. Aharonov and A. Casher, Phys. Rev. Lett. {\bf 53}, 319 (1984).

\bibitem{r3} B. Mashhoon, Phys. Rev. Lett. {\bf61}, 2639 (1988).


\bibitem{fur1} E. Passos, L. R. Ribeiro, C. Furtado, and J. R. Nascimento, Phys. Rev. A {\bf76}, 012113 (2007).

\bibitem{fur2} C. A. de Lima Ribeiro, C. Furtado and F. Moraes, Europhys. Lett. {\bf62}, 306 (2003).


\bibitem{bf1} K. Bakke {\it et al}, Phys. Rev. D {\bf78}, 064012 (2008); K. Bakke {\it et al}, Eur. Phys. J. C {\bf60}, 501 (2009).

\bibitem{bf8} K. Bakke and C. Furtado, Ann. Phys. (Berlin) {\bf522}, 447 (2010).

\bibitem{bf26} K. Bakke and C. Furtado, Phys. Lett. A {\bf375}, 3956 (2011).

\bibitem{anan} J. Anandan, Phys. Lett. A {\bf138}, 347 (1989).

\bibitem{whw} H. Wei, R. Han and X. Wei, Phys. Rev. Lett. {\bf75}, 2071 (1995).

\bibitem{hag} C. R. Hagen, Phys. Rev. Lett. {\bf77}, 1656 (1996).

\bibitem{whw2} H. Wei, X. Wei and R. Han, Phys. Rev. Lett. {\bf77}, 1657 (1996).

\bibitem{aud} J. Audretsch and V. D. Skarzhinsky, Phys. Lett. A {\bf241}, 7 (1998).

\bibitem{spa} G. Spavieri, Phys. Rev. Lett. {\bf81}, 1533 (1998).

\bibitem{spa1} M. Wilkens, Phys. Rev. Lett. {\bf81}, 1534 (1998).

\bibitem{spa2} G. Spavieri, Phys. Rev. Lett. {\bf82}, 3932 (1999).

\bibitem{spa3} G. Spavieri, Phys. Rev. A {\bf59}, 3194 (1999).

\bibitem{tka1} W. H. Heiser and J. A. Shercliff, J. Fluid Mech. {\bf22}, 701 (1985).

\bibitem{tka2} S. Y. Molokov and J. E. Allen, J. Phys. D {\bf25}, 393 (1992); S. Y. Molokov and J. E. Allen, J. Phys. D {\bf25}, 933 (1992).

\bibitem{tka3} C. Chryssomalakos, A. Franco, and A. Reyes-Coronado, Eur. J. Phys. {\bf25}, 489 (2004).

\bibitem{tka} V. M. Tkachuk, Phys. Rev. A {\bf62}, 052112 (2000).


\bibitem{lin} L. R. Ribeiro, C. Furtado, J. R. Nascimento, Phys. Lett. A {\bf348}, 135 (2006).

\bibitem{bf4} K. Bakke, L. R. Ribeiro, C. Furtado and J. R. Nascimento, Phys. Rev. D {\bf79}, 024008 (2009).

\bibitem{b} K. Bakke, Phys. Rev. A {\bf81}, 052117 (2010);	
						K. Bakke, Int. J. Mod. Phys. A {\bf26}, 4239 (2011).
						
\bibitem{bf5} K. Bakke and C. Furtado, Phys. Rev. A {\bf80}, 032106 (2009).
						
\bibitem{b4} K. Bakke, Ann. Phys. (Berlin) {\bf523}, 762 (2011).


\bibitem{belich} H. Belich, T. Costa-Soares, M. M. Ferreira Jr. and J. A. Helay\"el-Neto, Eur. Phys. J. C \textbf{41}, 421 (2005).

\bibitem{belich2} H. Belich, T. Costa-Soares, M. M. Ferreira Jr., J. A. Helay\"el-Neto and F. M. O. Moucherek, Phys. Rev. D \textbf{74}, 065009 (2006).

\bibitem{belich3} H. Belich, L. P. Collato, T. Costa-Soares, J. A. Helay\"el-Neto and M. T. D. Orlando, Eur. Phys. J. C \textbf{62}, 425 (2009).

\bibitem{bbs2} K. Bakke, H. Belich and E. O. Silva, Ann. Phys. (Berlin) {\bf523}, 910 (2011).

\bibitem{extra3} V. A. Kostelecky and S. Samuel, Phys. Rev. D {\bf39}, 683 (1989).

\bibitem{extra1} D. Mattingly, Living Rev. Relativity {\bf8}, 5 (2005).

\bibitem{extra2} V. A. Kostelecky, {\it CPT and Lorentz Symmetry} (World Scientific, Singapore, 2011).


\bibitem{col} D. Colladay and V. A. Kostelecký, Phys. Rev. D {\bf55}, 6760 (1997); D. Colladay and V. A. Kostelecký, Phys. Rev. D {\bf58}, 116002 (1998).

\bibitem{belic} H. Belich, T. Costa-Soares, M. M. Ferreira Jr., J. A. Helayel-Neto, M. T. D. Orlando, Phys. Lett. B {\bf639}, 675 (2006).

\bibitem{Hamilton} V. A. Kostelecky and C. D. Lane, J. Math. Phys. {\bf40}, 6245 (1999); R. Lehnert, J. Math. Phys. {\bf45}, 3399 (2004).

\bibitem{Manojr} M. M. Ferreira Jr. and F. M. O. Moucherek, Int. J. Mod. Phys. A {\bf21}, 6211 (2006); Nucl. Phys. A {\bf790}, 635 (2007); S. Chen, B. Wang, and R. Su, Class. Quant.Grav. {\bf23}, 7581 (2006); O. G. Kharlanov and V. Ch. Zhukovsky, J. Math. Phys. {\bf48}, 092302 (2007).

\bibitem{Nonmini} H. Belich, T. Costa-Soares, M. M. Ferreira Jr., J. A. Helay\"el-Neto and F. M. O. Moucherek, Phys. Rev. D {\bf74}, 065009 (2006).

\bibitem{proc} J. S. Diaz, Proceedings of the DPF-2011 Conference, Providence, RI, August 8-13, 2011; arXiv:1109.4620.



\bibitem{bbs} K. Bakke, H. Belich and E. O. Silva, J. Math. Phys. {\bf52}, 063505 (2011).

\bibitem{mano} H. Belich, E. O. Silva,  M. M. Ferreira Jr. and M. T. D. Orlando, Phys. Rev. D. {\bf83}, 125025 (2011).

\bibitem{greiner} W. Greiner, \textit{Relativistic Quantum Mechanics: Wave Equations, 3rd Edition} (Springer, Berlin, 2000).

\bibitem{dirac} P. A. M. Dirac, Proc. R. Soc. A {\bf133}, 60 (1931); M. V. Berry, Eur. J. Phys. {\bf1}, 240 (1980).



\bibitem{disp} G. Badurek, H. Weinfruter, R. G\"ahler, A. Kollmar, S. Wehinger, and A. Zeilinger, Phys. Rev. Lett. {\bf71}, 307 (1993).

\bibitem{disp2} M. Peshkin and H. J. Lipkin, Phys. Rev. Lett. {\bf74}, 2847 (1995).

\bibitem{disp3} M. Peshkin, Found. Phys. {\bf29}, 481 (1999).

\bibitem{abelian} S. Loyd, Quantum Inf. Proc. {\bf1}, 13 (2002).

\bibitem{schu} P. Schl\"uter, K.-H. Wietschorke and W. Greiner, J. Phys. A {\bf16}, 1999 (1983).

\bibitem{bd}  N. D. Birrell and P. C. W. Davies, \textit{Quantum Fields in Curved Space}, (Cambridge University Press, Cambridge, UK, 1982).

\bibitem{weinberg} S. Weinberg, {\it Gravitation and Cosmology: Principles and Aplications of the General Theory of Relativity} (IE-Wiley, New York, 1972).

\bibitem{mello} E. R. Bezerra de Mello, J. High Energy Phys. {\bf6}, 016 (2004).

\bibitem{b3} K. Bakke, J. Math. Phys. {\bf51}, 039516 (2010).

\bibitem{abra} M. Abramowitz and I. A. Stegum, \textit{Handbook of mathematical functions} (Dover Publications Inc., New York, 1965).

\bibitem{landau} L. D. Landau and E. M. Lifshitz, \textit{Quantum Mechanics, the nonrelativist theory, 3rd Ed.} (Pergamon, Oxford, 1977).






\end{thebibliography}
\end{document}